\documentclass[11pt,twoside]{article}
\usepackage{asp2014}

\aspSuppressVolSlug
\resetcounters

\begin{document}

\title{(I6.1) Why is the LSST Science Platform built on Kubernetes?}

\author{Christine~Banek,$^1$ Adam~Thornton,$^1$ Frossie~Economou,$^1$ Angelo~Fausti,$^1$ K.~Simon~Krughoff,$^1$ and Jonathan~Sick$^1$}
\affil{$^1$AURA/LSST, Tucson, AZ, USA; \email{cbanek@lsst.org}}

\paperauthor{Christine~Banek}{cbanek@lsst.org}{0000-0002-4337-4956}{AURA}{LSST}{Tucson}{AZ}{85721}{USA}
\paperauthor{Adam~Thornton}{athornton@lsst.org}{0000-0001-9342-6032}{AURA}{LSST}{Tucson}{AZ}{85721}{USA}
\paperauthor{Frossie~Economou}{frossie@lsst.org}{0000-0002-8333-7615}{AURA}{LSST}{Tucson}{AZ}{85721}{USA}
\paperauthor{Angelo~Fausti}{afausti@lsst.org}{0000-0002-8095-305X}{AURA}{LSST}{Tucson}{AZ}{85721}{USA}
\paperauthor{K.~Simon~Krughoff}{krughoff@lsst.org}{0000-0002-4410-7868}{AURA}{LSST}{Tucson}{AZ}{85721}{USA}
\paperauthor{Jonathan~Sick}{jsick@lsst.org}{0000-0003-3001-676X}{AURA}{LSST}{Tucson}{AZ}{85721}{USA}




\newcommand{\code}[1]{\texttt{#1}}

\begin{abstract}
LSST has chosen Kubernetes as the platform for deploying and
operating the LSST Science Platform.  We first present the
background reasoning behind this decision, including both
instrument-agnostic as well as LSST-specific requirements.
We then discuss the basic principles of Kubernetes and Helm, and how
they are used as the deployment base for the LSST Science Platform.
Furthermore, we provide an example of how an external group may
use these publicly available software resources to deploy their own
instance of the LSST Science Platform, and customize it
to their needs.  Finally, we discuss how more astronomy software
can follow these patterns to gain similar benefits.
\end{abstract}

\section{Introduction}

The Large Synoptic Survey Telescope \citep[LSST;][]{2019ApJ...873..111I}
dataset can not be easily stored,
copied, or analyzed, due to its large size.  Hundreds of machines will be
required, not only to store and process the data, but also to allow users
to run their own analysis next to the data.

The LSST Science Platform \citep[LSP;][]{LSE-319,LDM-554,LDM-542} provides next-to-data processing and
real-time interaction with the data.  By allowing scientists to run their own
Python code in Jupyter\footnote{https://jupyter.org/} notebooks in the LSP,
we run their code next to the data no matter where they are in the world.

\subsection{Without Reproducbility, No One Knows What You Found}

Over the 10 year survey, machines will experience hardware failures, and
will also be replaced with newer, faster, more efficient machines.
Software will be patched, updated, and reinstalled.  The system as a whole
must continue to function with some percentage of its nodes being unavailable.
As the system also needs to support user demand, the system
must also handle dynamic software loads.

These practical requirements call for an automated and reliable process for
deploying new systems or replicate environments from scratch in a reproducible
way.  If this process were not automated, the required manual process would unsustainably
drain developer resources to execute.  Any manual step is a chance for human error,
introducing subtle mistakes that could undermine scientific results.

At smaller scale, it is even more important to have reproducible software.  With
fewer machines, a larger fraction of computing resources could be unavailable due to
hardware or software failures, possibly to the point of unavailability of the entire
system.  However, if your machines take many years to fail, it may be discovered too
late that you cannot reproduce the system.
Long-lived machines also have greater risk of being manually changed by
well-meaning operators and developers attempting to hot-fix or diagnose issues.

Reproducing scientific results on an unstable foundation is fraught with risk.
While software will never be perfect, through careful management, results should
be reproducible (even if flawed), which will help discovery and resolution of
software bugs.  In an uncontrolled software environment, it can be hard to
differentiate between software bugs and the next Nobel Prize.

\subsection{The Quest for Reproducible Software}

Reproducing a working software environment is not trivial, due to
its natural complexity and interdependence between components.  These components
include astronomy applications, libraries and runtimes they depend upon, the
operating system, and other applications running concurrently on the same machine.
Since all these required applications cannot
typically run on one physical machine, applications interconnect over a network,
requiring services to assist inter-applications communication, such as DNS, proxies,
and load balancers.

Most software is designed to run under tightly controlled conditions, because
software is easier to write, test, and verify when you can rely on
well-specified conditions.  For example, supporting every version of every
operating system is a weighty, unnecessary burden that most software projects
should not attempt.

\subsection{Virtualization}

Using virtualization, a machine running one operating system can run a different
operating system inside a virtual machine (VM).  The physical machine and the VM
are isolated from each other.   VMs introduce a performance penalty for heavy CPU
processing or disk I/O.  Since each VM contains its own copy of the operating
system, libraries, and utilities, this overhead reduces the number of VMs a physical machine
can run.  Using VMs, it is possible, although costly, to provide complete isolation of
application instances by running each in an independent VM.

\subsection{Containerization}

Containers are a way of packaging applications and their environments as one self-contained unit.
Multiple containers can run on one host sharing one operating system kernel, rather than simultaneously
running many independent kernels in virtual machines.  This reduces the overhead of running many containers
versus running many VMs, while still providing much of the isolation of a VM.\footnote{Containers
achieve this by using various features in kernels such as namespacing and cgroups
to share the same kernel while isolating filesystems and process spaces.}  Containers
encourage further isolation between applications so that each application is run in its
own container.

Without the overhead of starting an operating system, containers can start very quickly,
typically within seconds.  Compare this to a VM, which may take minutes to be ready to handle
user requests, even when already installed and configured. This makes containers not only
easy to replicate but also quick to start on a new machine.

New containers are easy to build by extending existing containers.  Many operating
system containers already exist and are maintained by their respective organizations.
Starting from one of these operating system containers, additional software can be
installed and customized, files can be added, and commands can be run to create a
new purpose-built container hosting your software.  These steps are contained in a
\code{Dockerfile}, which can be committed to source control alongside the code.

Containers are immutable; on startup it always has the configuration it had when built.
Any changes inside a container after startup are lost upon restart.  For persistent storage,
volumes can be mounted inside a container, using the filesystem of the host
to allow for storage between container restarts.  While this may seem cumbersome,
by resetting the state, containers are more resilient to an unrecoverable state or corruption.

Containers are also easy to share on the Internet.  Docker Hub\footnote{https://hub.docker.com}
allows for users to publish container images, which consist of
a compressed \code{tar} file of the root filesystem.\footnote{For performance, this is split into
various layers, so that containers can share parts of the filesystem where possible} This allows
anyone to run your container on their local machine in minutes.  Containers images are published under
a name, along with tags that can represent different versions.\footnote{For example, lsst/application:latest}\textsuperscript{,}\footnote{
For an excellent tutorial on creating containers, refer to: https://docs.docker.com/get-started/
and https://docker-curriculum.com}

So, finally, we have a reliable method to replicate the installation of software,
at least on one machine.

\subsection{Kubernetes}

While containers are very useful, one container doesn't know about the other containers
on the system, and doesn't know anything about what might be running on other systems.
We need an orchestrator that can produce a complex system from individual containers
across multiple machines.  Kubernetes, or k8s for short, is an open-source system
built originally by Google to handle this exact problem.  Kubernetes allows for the
coordination and orchestration of containers across a cluster of machines.

\section{Kubernetes: Core Concepts}

K8s is a declarative system - instead of running
a sequence of commands, you declare what you want the final state of the system to be.
K8s will perform the required actions to create and maintain that state over time.
K8s defines this state as a set of resources with configurable parameters, each represented by a YAML document.
These resources are created by using the \code{kubectl} command line utility.

The main way of creating groups of containers is by using the \code{deployment} resource.\footnote{
Technically, deployments act on pods, which are groups of heterogeneous containers that should be
scheduled on the same host.  But many times, a pod has one container.}
A k8s \code{deployment} resource instructs Kubernetes to schedule a number of containers to run,
to check the health of those containers,
and take actions to ensure they are kept running and available.  If one container exits, doesn't respond
to a health check, or the machine hosting it becomes unavailable, that container is then scheduled
on a different machine automatically.\footnote{
Here is a hands-on tutorial that goes command by command through this process:
https://kubernetes.io/docs/tutorials/hello-minikube/\#create-a-deployment}

A k8s \code{deployment} also allows a number of replicas to be set.  To run more copies of your application to
scale up, use the \code{kubectl} tool to change the number of replicas for
a \code{deployment}.  If the number of running copies falls below the number of desired replicas, new
containers are scheduled and created.  If there are more containers than desired, containers are terminated
until the count reaches the desired number of replicas.

At this point, you have a container in a \code{deployment}, but since you don't know
where it is running, it is hard to access it externally.
Moreover, since a container might be moved or restarted on a different machine, you need to discover
where the container is currently running.  By creating
a \code{service} resource, you can internally contact containers using a DNS name.
This DNS name is defined by the \code{service} name, and will load-balance
requests among all the running containers.

By default, containers running in k8s are connected to a private internal network shared by all
the k8s nodes.  To expose a service to the external network, you will need to create an \code{ingress}
resource.  \code{ingress} resources map public DNS names and HTTP(S) paths
to backend k8s \code{service} instances.  These requests are routed through an ingress
controller, such as the \code{nginx-ingress} controller.  Each ingress rule can have
individualized configuration of nginx features such as a timeouts and URL rewriting.
By using \code{ingress} resources you can avoid having individual nginx instances for each service.
It is also easy to host multiple web services or applications behind one public DNS name,
routing requests to their respective container via the URL, creating a typical reverse proxy.

Permanent storage is required for many applications, and since k8s may move containers
between machines, being able to utilize the same storage across hosts is an important
consideration.  K8s provides a way to allocate and attach storage to containers.
The \code{persistent volume} (\code{PV}) defines the type of storage to use, and how
much space to allocate.  Each \code{PV} can then be mounted by creating a \code{persistent
volume claim} (\code{PVC}), which is then referenced in the \code{deployment} YAML.  In this way, your
container can mount a \code{PVC} on a specific path inside the container.
However, unlike the container filesystem, the state of the \code{PV} will be kept
between container restarts.

Configuration of containers is another important consideration.  K8s provides several ways of
managing configuration.  The simplest is the container environment.  In the \code{deployment}
resource, each container is defined, including the UNIX environment variables to set.
Due to the ubiquitous use of UNIX environment variables for controlling programs, intended for
containers or not, this can be a very simple and effective method for configuring containers.

Another type of configuration is the \code{configmap}.  K8s \code{configmap}s are typically
groups of files each with a filename.  Like mounting a \code{PV}, a \code{configmap}
can be mounted into a container, presenting the files in the configmap as normal files.
In this way, configuration files can be injected into any container,
without having to be present in the container initially.

One last type of configuration to mention is the k8s \code{secret}.  K8s \code{secret}s can
be used to hold sensitive configuration files of any type, passwords, or certificates.  Special
care is taken to not expose the values of these secrets, which are shown for
environment variables and configmaps.  By using \code{secret}s it is easier to tell which
parts of the configuration are secret (and likely instance specific) and which are general
configuration.  \code{secret}s can be mounted like a volume into the containers
where they are used or as environment variables.

\section{Helm}

K8s can be intimidating, and each application can have many parts.  For those who only
want to install and run software, rather than develop it, looking at the raw
k8s YAML documents might be too confusing and create a high barrier to entry.

Since the birth of k8s, many have tried to solve this problem, and one of these
solutions is Helm.\footnote{https://helm.sh}  Helm is a package manager for k8s, allowing
for installation of all the parts of an application in one command.
Most importantly, Helm allows for customization of the installed software.

Helm's packages are called charts.  Each chart is a \code{tar} archive containing a set of
k8s YAML resources to apply to the cluster.  These YAML files are first run through
a templating engine.  Templating allows for user provided values to be injected
into the correct place in the YAML resources sent to k8s.
The values to apply to the template are provided via a user specific YAML document.
This allows for the user to build a simpler YAML document with only the configuration
values they want to apply, without having to have deep k8s knowledge.  In this way,
users can now run software on k8s without needing to know all the details of developing
and configuring that software.

Helm is one of the leading software packaging schemes for k8s, and many charts
already exist for commonly used open-source software.\footnote{https://github.com/helm/charts
is the repository for official Helm charts}

\section{Using Helm for Installing the LSP}

To help everyone be able to install the LSP, there is now a repository of Helm
charts available at https://github.com/lsst-sqre/charts.  Each of these components
of the LSP has its own chart that anyone can install:

\begin{itemize}

\item landing-page: A simple HTML landing page for the LSP with links to help users
discover the different components of the LSP.  This also provides a simple message of
the day to users, allowing operators and administrators to present relevant
messages about system upgrades or downtime.

\item fileserver: A simple NFSv4 fileserver that provides shared storage for
home directories and data.  This is mounted by the Nublado JupyterLab containers,
and allows for persistent storage between logins, as well as a place to centrally
store data to process.  For those with an existing NFS fileserver in the environment,
this won't be used, instead Nublado can be configured to point to your existing
fileserver.

\item nublado: LSST's JupyterHub environment.  Nublado allows for users to spawn
JupyterLab containers in k8s.  Users can then run Jupyter notebooks and perform interactive
data analysis next to the data.  This chart provides a number of features not present
in stock JupyterHub, including a page to allow users to choose between different versions
of their Jupyter environment.

\item firefly: Firefly \citep{2013ASPC..475..315R} is one of the ways users can interact with image and catalog
data.  Firefly is a web portal that allows for quick preliminary
investigation using the browser, and is integrated into the Jupyter
environment for interactive analysis with notebooks.

\item cadc-tap: The cadc-tap chart installs an IVOA TAP \citep[Table Access Protocol;][]{2010ivoa.spec.0327D}
server into k8s, allowing users to query catalogs through Nublado, Firefly,
or external clients.  This chart also sets up simple database backends
in k8s to provide TAP\_SCHEMA and a sample dataset for the TAP service to serve.
The server that this chart installs was written by the Canadian Astronomy Data Centre,
and the code is publicly available on GitHub.\footnote{https://github.com/opencadc/tap}

\end{itemize}

Convenience scripts with a working configuration for Google are provided here:
https://github.com/lsst-sqre/lsp-deploy\footnote{Given that the instructions
may change over time, I won't go through the exact
instructions here, but instead explain what is happening.  The authoritative instructions
are in the lsp-deploy repository along side the code, so they can be updated together
and remain relevant over time.}

There are a few stages to installing the LSP, which we will cover next.

\subsection{Creating a Kubernetes Cluster}

Either you have an existing k8s cluster, or you will create one.  It is very
easy to create a k8s cluster on Google using Google Kubernetes Engine (GKE).
While this can be done on the command line, it can be done very easily
using the Google Cloud Console.\footnote{https://cloud.google.com/kubernetes-engine/docs/how-to/creating-a-cluster
provides a good walkthrough of creating a cluster using both the command line
and the web portal.}

In general, you want your k8s cluster to contain at least 3 nodes, with each node
having at least 2, if not 4, cores.  Also increase the host disk from
the default 100GB to 200GB to store the container images.

Once you've created your cluster, you can connect it to your local \code{kubectl}
command by clicking the 'Connect' button, and running the command line provided.

\subsection{Setting up Helm}

Helm has a local component and a k8s component that both need to be installed.
The \code{install\_tiller.sh} script provides a simple wrapper to set up the
server side Helm components.
By default, this isn't particularly secure, so you should look into the Helm
documentation to create a more secure deployment for permanent environments.
However it will serve to install the various LSP components on a test instance.
You can tell if Helm is working properly by running \code{helm ls}, and verifying
that nothing is installed, but no error is generated.

\subsection{Preinstallation Steps}

The next step is installing and configuring the \code{nginx-ingress} controller
using the Helm chart.  These steps are provided in \code{install\_ingress.sh}.
This script creates a k8s \code{secret} holding the SSL certificate for the cluster, installs the Helm
chart, and configures nginx to use the created \code{secret} for the certificate.\footnote{
If you don't already have a current SSL certificate, you can create one for
free using letsencrypt.org at https://letsencrypt.org/getting-started/}

Next is a manual step of setting the DNS record for your new cluster.
Run the \code{public\_ip.sh} script to retrieve the external IP address for your
cluster, and assign it to a DNS A record that matches the SSL certificate.

At this point, if everything went well, you should be able to go to your
DNS name over HTTPS, have it resolve, and be presented with a 404 that says
\code{default backend}.

Nublado requires GitHub authentication to verify people can log in.  Follow
the GitHub instructions\footnote{
https://developer.github.com/apps/building-oauth-apps/authorizing-oauth-apps/
} for creating an OAuth application, and put the resulting client ID and secret in the
\code{nublado-values.yaml} file.

\subsection{Installing LSP Charts with Helm}

Now let's examine the \code{install\_lsp.sh} script.  This script adds the LSP
chart repository to Helm's search path, then uses \code{helm install} to install
each chart in turn.  After you run this script, you should be able to run \code{helm ls}
and see that multiple charts are now installed along with their versions.  If you
now return to your DNS name in the browser, you should be presented with the
LSP Landing Page.  This provides links to the Nublado notebook environment and
Firefly.

\section{Customizing the Landing Page}

Now that you have a running LSP, it is easy to inspect it, use it, and see
how it works in practice.  One of the first things you may want to do is
customize the landing page.  After all, it would be confusing to present
yourself as the LSST Science Platform if you are not.  Here we will go
through a few different examples using different types of customization
to demonstrate how easy it is to tailor the LSP using Helm.

\subsection{Changing the MOTD}

One simple thing you will want to do is change the message of the day,
which is presented in a box on the landing page. The message of the day
comes from a markdown document whose URL is passed into the container
as an environment variable.  We will use Helm to allow us to customize that parameter
to point to a new markdown document.  Edit \code{landing-page-values.yaml}
and replace the \code{motd\_url} URL with the URL of a markdown document
of your choice. Now when the LSP is installed, the message of the day should display
the markdown document you specified.

\subsection{Changing the Container Image}

Instead of just changing the MOTD, if you want to use a different HTML landing page,
the easiest way is to create a new container containing the desired web page.  You can
examine how the docker container for the landing page is created by looking at the
Dockerfile in the landing page repository.\footnote{
https://github.com/lsst-dm/lsp-landing-page/blob/master/Dockerfile}
After the new container image is published to \code{hub.docker.com}, edit
\code{landing-page-values.yaml} to use that container image name instead of the default.

\subsection{Fixing Bugs and Suggesting Chart Improvements}

Bugs are the nature of software development, and Helm charts are no different.  If you
find a bug, please create an issue on the chart repo\footnote{https://github.com/lsst-sqre/charts}
on GitHub.  This will allow others to see there may be a problem, as well as notify the
developers that something may need to be fixed.

If you already know what the required fix is, please submit a pull request (PR)
to the repository with your changes.\footnote{
https://help.github.com/en/articles/creating-a-pull-request}

If you have new features you wish to add to an existing chart, such as new ways of overriding
particular deployment values, you can also submit a PR containing that change.  It helps
if the changes are backward compatible, but sometimes breaking compatibility is required.

\section{Publishing Astronomy Software with Helm}

The Helm charts listed in section 4 are not intended to be a complete or authoritative
list, but a start.  I hope that after seeing the usefulness of being able
to install another institution's science platform, your institution will also want to
use k8s and Helm to share software with the rest of the astronomy community.

The Helm repository framework allows for anyone to pull charts from multiple chart
repository sources without having to have a centralized store, gate-keeper, or process.  It is
simple to create your own repository on GitHub to store your own Helm charts.
This allows for your Helm charts to be developed under whatever software development
process or timeline your institution recommends or requires.\footnote{
https://helm.sh/docs/chart\_repository/ is an excellent resource for using existing
services such as GitHub Pages and others to host your helm charts for free.}

If you are creating a new chart for a new software service, there are many great
blog posts and a wealth of helpful documentation on the Helm website.\footnote{
https://docs.bitnami.com/kubernetes/how-to/create-your-first-helm-chart/
is a very thorough walkthrough of an example}

\section{Conclusion}

The field of astronomy will be ever more reliant on software to produce its results.
Being able to reproducibly develop, deploy, configure, and operate software are
essential initial steps to relying on that software to produce scientific
results that can be replicated, reproduced, and trusted.

As astronomy software becomes more complex and intertwined with other big
data science fields, such as computer science, biology, and physics, it will
become necessary to share software within and between these groups.  Writing software that
is of high quality and scales to these problems requires specialized talent,
and a large time investment.  It makes logical, scientific, and financial
sense for disparate fields to collaborate on software where possible,
rather than each institution developing its own specialized software
to solve similar problems.

In the era of big data and multi-dataset astronomy, it is practically required
for institutions that mirror an instrument's dataset to also provide local mirrors
of software services and tools to operate on those datasets.  Given the specialized
nature of each instrument, it will be increasingly important
to allow astronomers to work with multiple datasets co-located in the same datacenter.
These multi-dataset datacenters will require a standard
way of installing and updating software provided by the instrument creators and software
developers with limited assistance. Kubernetes and Helm provide one way of accomplishing
this goal using software practices and tools that are already standard in the software
industry.

By bringing in tools and practices from the software industry, we can also
leverage the open source community (as well as companies who participate in
open source) as a force multiplier for astronomers to accomplish
more than they ever could before. Using these industry standard tools that
developers know (and love), it could also be a boon to recruiting talented
software developers to astronomy.  As astronomers become more familiar with
these new tools, and by sharing them across multiple instruments, that
familiarity can help increase the efficiency of analysis and scientific
output, as well as reduce the time to verify and replicate results.

\section{Acknowledgements}

This material is based upon work supported in part by the National Science
Foundation through Cooperative Agreement 1258333 managed by the Association
of Universities for Research in Astronomy (AURA), and the Department of Energy
under Contract No. DE-AC02-76SF00515 with the SLAC National Accelerator Laboratory.
Additional LSST funding comes from private donations, grants to universities, and
in-kind support from LSSTC Institutional Members.

We thank Matias Carrasco Kind, Tim Jenness, and Kian-Tat Lim for reviewing
this manuscript.

\bibliography{I6-1}


\end{document}